\documentclass{jfm}
\usepackage{graphicx}
\usepackage{epstopdf, epsfig}
\usepackage{amsmath}
\usepackage{cases}
\usepackage{color}
\usepackage{hyperref}
\usepackage{amssymb}
\usepackage{graphics}
\usepackage{subfigure}
\usepackage{textcomp}
\usepackage{bm}
\usepackage{url}
\usepackage{tikz}
\allowdisplaybreaks[1]

\makeatletter
  \newcommand\figcaption{\def\@captype{figure}\caption}
  \newcommand\tabcaption{\def\@captype{table}\caption}
\makeatother

\shorttitle{Explicit series solution of a capillary flow}
\shortauthor{X.  Zhong, B. Sun and S.  Liao}

\title{A singularity-free analytic solution of rise dynamics of a liquid in a vertical cylindrical capillary}

\author{Xiaoxu Zhong\aff{3}, Bohua Sun\aff{4}
  \and Shijun Liao\aff{1,2,3}
  \corresp{\email{sjliao@sjtu.edu.cn}},
  }

\affiliation{\aff{1} State Key Laboratory of Ocean Engineering, Shanghai 200240, China
\aff{2} Collaborative Innovative Center for Advanced Ship and Deep-Sea Exploration\\ Shanghai 200240, China
\aff{3} School of Naval Architecture, Ocean and Civil Engineering\\
Shanghai Jiao Tong University, Shanghai 200240, China\\
\aff{4} Faculty of Engineering\\ Cape Peninsula University of Technology, Cape Town 7535, South Africa
}

\begin{document}

\maketitle

\begin{abstract}

Capillary driven flow is a famous problem in fluid dynamics which dates back to L\'{e}onardo da Vinci \citep{Guillaume2016}.  In this paper, we apply an analytic approximation method for highly nonlinear problems, namely the homotopy analysis method (HAM), to a model of the meniscus movement in a uniform vertical circular tube.  Convergent explicit series solution is successfully obtained.  Our results agree well with the numerical results given by the symbolic computing software Mathematica using six-order Runge-Kutta methods.  More importantly, our analytic solution is valid in the whole region of physical parameters, and therefore can predict whether the path of liquid is monotonic or oscillatory.  This kind of solution, to the best knowledge of the authors, has never been reported in the past, which might greatly deepen our understandings about capillarity.
\end{abstract}


\section{Introduction}

Capillarity is one of the most usual phenomenon in nature, which has received wide attention from researchers \citep{Bosanquet1923On, Sparrow1964Flow, Batchelor1967, BLAKE1969Kinetics, FISHER1979An, JEJE1979Rates, JOOS1990The,  ICHIKAWA1994Interface, Duarte1996The, Romero1996Flow, Weislogel1998Capillary, David1999Rebounds, HAMRAOUI2000Can, Zhmud2000Dynamics, Hall2007Rising, Higuera2008Capillar, FRIES2009Dimensionless, MULLINS2012Capillary, CAI2012An, Bush2013Interfacial, Shardt2014Inertial, Fazio2014An, Lade2017Dynamics}.  The first quantitative treatment of capillary action was attained by \citet{Young1805} and \citet{Laplace1805}, who derived the famous Young-Laplace equation of capillary action that describes the capillary pressure difference sustained across the interface between two static fluids due to the surface tension.  After that, the boundary conditions governing capillary action was given by \citet{Gauss1830}.

\begin{figure}
  \centerline{\includegraphics[width=3.5in]{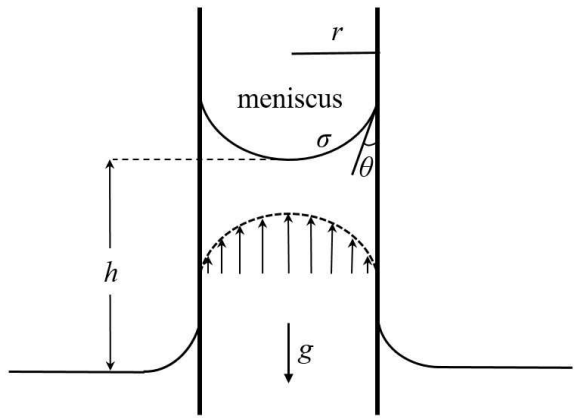}}
  \caption{Rise dynamics of a liquid in a vertical circular tube.}
\label{figure:Model}
\end{figure}

The dynamics behavior of meniscus is primarily determined by the balance between the inertia of the fluid, the capillary driving force, the weight of the liquid and the resisting viscous forces.   The model characterizing rise dynamics of a liquid in a vertical circular tube is as shown in Fig.~\ref{figure:Model}, where $r$ denotes the radius of the vertical circular tube and $g$ is the acceleration due to gravity, $\theta$ represents the dynamic contact angle between the surface of the liquid and the wall of tube, $\sigma$ denotes the surface tension coefficient, respectively.  The first comprehensive investigation of the capillary rise dynamics dates back to \citet{Bell1906The}, \citet{Lucas1918} and \citet{Washburn1921}, who considered the capillary flow in a cylindrical tube of radius $r$ and presented the rate of penetration as
\begin{equation}
\frac{\textrm{d}h}{\textrm{d}t}=\frac{P(r^2+4\epsilon r)}{8\mu h}, \label{1:equ:full:washburn}
\end{equation}
where $t$ is the time for a liquid of dynamic viscosity $\mu$ and slip coefficient $\epsilon$ to penetrate a distance $h$ into the capillary under the driven pressure $P$.  However, in the initial stage, i.e., $t\rightarrow 0$,  $h\rightarrow 0$, $\textrm{d}h/\textrm{d}t\rightarrow \infty$ can be derived from Eq. (\ref{1:equ:full:washburn}), which is obviously contrary to natural phenomenon.    Since then, a set of nonlinear differential equations were presented basing on different considerations of inertial force, entrance effect and dynamic contact angle, etc.

\citet{Brittin1946Liquid} assumed that the forces acting on the liquid in an accelerating state are the same as in a steady state, and then proposed an equation describing the dynamics of capillary rise:
\begin{equation}
h\frac{\textrm{d}^2 h}{\textrm{d} t^2}+\frac{5}{4}\left(\frac{\textrm{d}h}{\textrm{d}t} \right)^2+\frac{8\mu}{\rho r^2}h\frac{\textrm{d}h}{\textrm{d}t}-\frac{2\sigma \cos \theta}{\rho r}+gh=0, \label{1:equ:Brittin}
\end{equation}
in which $\rho$ denotes the density of liquid. By expressing $h(t)$ in a double Dirichlet series, \citet{Brittin1946Liquid} obtained approximation solutions which agree with experimental data.  After that, \citet{SZEKELY1971The} took the energy dissipation into account and proposed a more rigorous equation which successfully removes the singularity in the Washburn equation (\ref{1:equ:full:washburn} and \ref{1:equ:Brittin}):
\begin{equation}
\left(h+\frac{7r}{6} \right)\frac{\textrm{d}^2 h}{\textrm{d} t^2}+1.225\left(\frac{\textrm{d}h}{\textrm{d}t} \right)^2+\frac{8\mu}{\rho r^2}h\frac{\textrm{d}h}{\textrm{d}t}-\frac{2\sigma \cos \theta}{\rho r}+gh=0. \label{1:equ:Szekely}
\end{equation}
However, their treatment of dissipative effect refers to high Reynolds number, whereas most wetting phenomena in capillaries occur at low Reynolds number \citep{Levine1976A}.  So, an improved theory was further proposed by \citet{Levine1976A}:
\begin{equation}
\left(h+\frac{37r}{36} \right)\frac{\textrm{d}^2 h}{\textrm{d} t^2}-\frac{2\sigma}{\rho r}+\frac{8\mu}{\rho r^2}h\frac{\textrm{d}h}{\textrm{d}t}+gh+\frac{1}{2}\left[\frac{4 \eta}{\rho r}\frac{\textrm{d}h}{\textrm{d}t}+\frac{7}{3} \left(\frac{\textrm{d} h}{\textrm{d} t}\right)^2   \right]=0. \label{1:equ:Levine}
\end{equation}
\citet{XIAO2006A} proposed a generalized model for capillary flows in parallel plates and tubes, and then solved this model by expressing unknown function into a double Dirichlet series as well, which is similar to \citet{Brittin1946Liquid}.  Subsequently, \citet{MAGGI2012Multiphase} considered the coupled effect of liquid-gas interactions and proposed a governing equation of two-phase flow in capillaries.  Under some assumptions \citep{MAGGI2012Multiphase}, the governing equation reduces to
\begin{equation}
\left(h+\frac{7r}{6} \right)\frac{\textrm{d}^2 h}{\textrm{d} t^2}+\frac{1}{2}\left(\frac{\textrm{d}h}{\textrm{d}t} \right)^2+\frac{8\mu}{\rho r^2}h\frac{\textrm{d}h}{\textrm{d}t}-\frac{2\sigma \cos \theta}{\rho r}+gh=0. \label{1:equ:Maggi}
\end{equation}
Note that there is a slightly difference between Eq.~(\ref{1:equ:Szekely}) and (\ref{1:equ:Maggi}), this is because  \citet{MAGGI2012Multiphase} did not include dissipation due to the `vena contracta' effect at the capillary entrance.  The same equation (\ref{1:equ:Maggi}) can also be found in \citep{Bush2013Interfacial}.

The process of capillary rise in a circular cylindrical tube can be divided into several stages \citep{Stange2003Capillary, Fries2010Capillary}.  At the beginning, the viscous effect is insignificant so that the first stage can be regarded as purely dominated by inertial forces of the initially accelerated liquid mass.  By neglecting the viscous and the gravity term, $h\propto t^2$ is found in this stage \citep{DREYER1994Capillary, Stange2003Capillary}.  With the influence of viscosity increases, the inertia dominated flow gradually evolves into viscous flow.  Detailed analysis of the transition from inertial to viscous flow can be found in the papers of \citet{Stange2003Capillary}, \citet{FRIES2008The} and \citet{Fries2010Capillary}.  Subsequently, the effect of inertia vanishes and the flow becomes purely viscous.  By neglecting both the influence of inertia and gravity, $h\propto \sqrt{t}$ can be derived in the purely viscous time stage \citep{Lucas1918, Washburn1921}.  After that, the purely viscous flow enters in the viscous and gravitational time stage.  the analytic solutions in this stage (neglecting inertia) in implicit and explicit forms are given by \citet{Washburn1921} and \citet{FRIES2008An} respectively.
At end, the equilibrium height or Jurin height $2\sigma \cos\theta/(\rho g r)$ \citep{Jurin1717}, where meniscus stops, is determined by the balance of gravity and surface tension force. It is worth mentioning that the oscillatory path of liquid in a capillary was also found in experiments \citep{Quere1997Inertial, Stange2003Capillary}.  For a defined liquid and capillary tube, a critical value for capillary radius was suggested by \citep{HAMRAOUI2002Analytical, MASOODI2013Dynamics}, below which the oscillation disappears.  Besides, the imbibition dynamics depends on the shape of free surface as well\citep{Tani2015Towards}.  A universal law $h\propto t^{1/3}$ was found when liquid rises into corners of different geometries \citep{ponomarenko2011A} or into short pillars with rounded edges, but dynamics is found to follow $h\propto \sqrt{t}$ when liquid rises into long pillars with sharp edges.  In addition, \citet{SIEBOLD2000Effect} found that the contact angle depends strongly on the rising velocity.

Various experiments have been done to observe the behaviors of capillary action and to confirm the validity of theories. But unfortunately most of them fail to cover the complete process.  The first experiments \citep{Bell1906The, Washburn1921, Elmo1945Data} were performed under normal gravity condition with small tube diameter, however only the $h\propto \sqrt{t}$ behavior was observed.  After that, \citet{Siegel1961Transient} carried out free fall experiments in a low-gravity (including zero-gravity) environment with different tube diameters, and successfully observed the $h\propto t$ behavior.  Subsequently, \citet{DREYER1994Capillary} performed a drop tower experiment, which verifies their theory that the capillary rise process is divided into three successive stages with at the beginning $h\propto t^2$, then $h\propto t$, and ultimately $h\propto \sqrt{t}$.  These three regions later were verified by \citet{Stange2003Capillary} who performed experiments under micro-gravity environment by using a $4.7\textrm{s}$ drop tower.  Moreover, \citet{Stange2003Capillary} gave a detailed analysis of why many experiments can only observe part of behavior.  \citet{Stange2003Capillary} considered two characteristic time scales $t_1\propto \sqrt{\rho r^3/\sigma}$, $t_2\propto r^2/\nu$ and the Ohnesorge number $\textrm{Oh}=\sqrt{\rho\nu^2/(\sigma d)}$ and the Bond number $\textrm{Bo}=\rho g r^2/\sigma$, and then proposed that the three phases of capillary rise process are separated by $t_1$ and $t_2$ which are determined by $\textrm{Oh}$ and the inertia of the liquid.  In more detail, if $t_1$ and $t_2$ are small (corresponding to large $\textrm{Oh}$ which often happens in ground experiment), only $h\propto \sqrt{t}$ behavior can be easily observed; but if wider tubes were used (corresponding to smaller $\textrm{Oh}$ which mainly happens in micro-gravity experiments, i.e., $\textrm{Bo}\rightarrow 0$ ), then $h\propto t$ and $h\propto t^2$ behavior can be observed.

Here, it is worth mentioning that, up to now, above equations characterizing the capillary action are mainly solved by numerical methods. For instance, a Lattice-Boltzmann method based on field mediator was proposed by \citet{WOLF2010Capillary} to simulate the capillary rise between parallel plates.  Sometimes, researchers will neglect certain individual terms in different flow regimes so as to obtain analytic solutions \citep{HAMRAOUI2002Analytical, FRIES2008An}.  Besides, although a double Dirichlet series $h=\sum_{m=0}^{\infty}\sum_{n=0}^{\infty}a_{m,n}\textrm{exp}[(mr_1+nr_2)t]$ was employed by \citet{Brittin1946Liquid} and \citet{XIAO2006A} and also their obtained series approximations agree well with numerical results as well, however they only considered the real $r_1$ and $r_2$.  In other words, their series approximations cannot describe the oscillatory path of liquid in a capillary. A Taylor's series solution and a perturbation method with perturbation quantity $\textrm{Bo}=\rho g r^2/\sigma$ were ever obtained by \citet{Sun2017On} for this problem, but unfortunately, \citet{Sun2017On} found that the rate of convergence of both solutions is extremely slow and thus useless for practical application.  Although \citet{Sun2017On} proposed an analytic approximate solution to improve the convergence, the solution could not properly predict the oscillatory path of liquid in a capillary found in experiments \citep{Quere1997Inertial, Stange2003Capillary}. Therefore, convergent analytic solutions of this problem, which are valid in the whole region and for different pathes (monotonic or oscillatory) of liquid in a capillary approaching the equilibrium position, have never been obtained until now, to be best knowledge of authors.

In this paper, an analytic approximation method, namely the homotopy analysis method (HAM), is employed for this problem. Unlike perturbation methods, the HAM is independent of \emph{any} small/large physical parameter. Especially, the HAM provides us great freedom to choose appropriate basis function, auxiliary linear operator and a so-called ``convergence-control parameter" $c_0$.  By means of these freedom, something new have been gained by the HAM: (1) the steady-state resonant waves were first predicted by the HAM in theory \citep{LIAO20111274, xu2012JFM, Liu2014JFM} and then confirmed experimentally in the laboratory \citep{Liu2015JFM}; (2) accurate results (including the wave profile enclosing sharp pointed angle $120^\circ$) of the limiting Stokes' wave in extremely shallow water were obtained by the HAM for the first time \citep{Zhong2017On}; (3) a uniform valid analytic solution of two-dimensional viscous flow over a semi-infinite flat plate was successfully given by the HAM for the first time \citep{Liao1999A}.  It is these freedom that distinguish the HAM from other analytic methods \citep{Liaobook, liaobook2, KV2008, Mastroberardino2011Homotopy, Kimiaeifar2011Application, KV2012, Duarte2015, Liao2016JFM, Zhong2017, Zhong2018Analytic, Liu2018Finite, Liu2018Mass}.  By means of the HAM, convergent explicit analytic solution of mathematical model (\ref{1:equ:Maggi}) is successfully obtained.  In addition, our results agree well with numerical results given by the symbolic computing software Mathematica using six-order Runge-Kutta methods.  More importantly, our analytic solution is valid for all cases, including the oscillatory action.  All of these show the power and potential of the HAM for complicated nonlinear equations.

To render the study self-contained, the paper is organized as follows. Following an introduction,  asymptotic property at $t\rightarrow +\infty$ of the rise dynamics is given in \S~2. Procedures of the HAM for the capillary action are presented in \S~3. Four cases including the monotonic and oscillatory path are considered in \S~4. Concluding remarks are given in \S~5.

\section{Asymptotic property}
We consider the singularity-free model proposed by \citet{MAGGI2012Multiphase}:
\begin{equation}
\left(h+\frac{7r}{6} \right)\frac{\textrm{d}^2 h}{\textrm{d} t^2}+\frac{1}{2}\left(\frac{\textrm{d}h}{\textrm{d}t} \right)^2+\frac{8\mu}{\rho r^2}h\frac{\textrm{d}h}{\textrm{d}t}-\frac{2\sigma \cos \theta}{\rho r}+gh=0. \label{2:equ:governing}
\end{equation}
subjects to the boundary conditions
\begin{equation}
h(0)=h'(0)=0. \label{2:equ:boundary}
\end{equation}
The dimensionless form of Eqs.~(\ref{2:equ:governing}) and (\ref{2:equ:boundary}) read
\begin{equation}
{\cal N}\big{[}z(\tau)\big{]}=A\;\frac{\textrm{d}^2 z}{\textrm{d} \tau^2}+z-1+z\;\frac{\textrm{d}^2 z}{\textrm{d} \tau^2}+\frac{1}{2}\left( \frac{\textrm{d} z}{\textrm{d} \tau}\right)^2+8B\;z\;\frac{\textrm{d} z}{\textrm{d} \tau}=0, \label{2:equ:dimensionless:governing}
\end{equation}
where $\cal N$ denotes a nonlinear operator, subjects to boundary conditions
\begin{equation}
z(0)=z'(0)=0. \label{2:equ:dimensionless:boundary}
\end{equation}
with the definition
\begin{eqnarray}
\left\{
\begin{split}
& A=\frac{7\textrm{Bo}}{12\cos \theta},\qquad B=\sqrt{\frac{2\cos \theta}{\textrm{Bo}\;\textrm{Ga}}}, \qquad z=\frac{h}{H}, \qquad \tau=\frac{t}{T}, \\
&\textrm{Bo}=\frac{\rho g r^2}{\sigma},   \qquad \textrm{Ga}=\frac{\rho^2 g r^3}{\mu^2},\qquad H=\frac{2\sigma\cos\theta}{\rho g r}, \qquad T=\sqrt{\frac{H}{g}}. \label{2:equ:definition}
\end{split}
\right.
\end{eqnarray}

Before we apply the HAM, let us firstly analyze the asymptotic property of $z(\tau)$ as $\tau\rightarrow +\infty$.  Let
\begin{equation}
z(\tau)=1+\varepsilon f(\tau),  \label{2:equ:asymptotic:analysis}
\end{equation}
in which $\varepsilon$ is a small quantity.  Then substituting (\ref{2:equ:asymptotic:analysis}) into Eq.~(\ref{2:equ:dimensionless:governing}) and let the coefficient of $\varepsilon$ be zero, we have
\begin{equation}
(A+1)\frac{\textrm{d}^2f(\tau)}{\textrm{d}\tau^2}+8B\frac{\textrm{d}f(\tau)}{\textrm{d}\tau}+f(\tau)=0.
\label{2:equ:asymptotic:coefficient}
\end{equation}
The solution of (\ref{2:equ:asymptotic:coefficient}) reads
\begin{equation}
f(\tau)=a_1 \;\textrm{e}^{\eta_1 \tau}+a_2\;\textrm{e}^{\eta_2 \tau},
\end{equation}
in which $a_1$ and $a_2$ are constant coefficients, and
\begin{equation}
\eta_1=\frac{-4B+\sqrt{16B^2-A-1}}{A+1},\qquad \eta_2=\frac{-4B-\sqrt{16B^2-A-1}}{A+1}. \label{2:eta1:eta2}
\end{equation}
Obviously, $\eta_1$ and $\eta_2$ are real when $\sqrt{16B^2-A-1}\geq 0$, this means that $f(\tau)$ approaches 1 exponentially as $\tau$ tends to infinite.  But if $\sqrt{16B^2-A-1}< 0$, the path of $f(\tau)$ approaching 1 is oscillatory since there exist imaginary numbers in $\eta_1$ and $\eta_2$.  According to these information, we introduce the following transformation
\begin{equation}
u=\textrm{e}^{\eta_1 \tau},\quad\quad u_c=\textrm{e}^{\eta_2 \tau}, \qquad w(u)=1-z(\tau). \label{2:equ:transformation}
\end{equation}
Then it is easy to know
\begin{equation}
\frac{\textrm{d}z}{\textrm{d}\tau}=-\eta_1 \;u\; \frac{\textrm{d}w}{\textrm{d}u},\qquad
\frac{\textrm{d}^2z}{\textrm{d}\tau^2}=-\eta_1^2 \left[ u \;\frac{\textrm{d}w}{\textrm{d}u}+u^2 \; \frac{\textrm{d}^2w}{\textrm{d}u^2} \right], \label{2:equ:transformation:derive}
\end{equation}
in addition,
\begin{eqnarray}
\left\{
\begin{split}
&\frac{\textrm{d}u_c}{\textrm{d}u}=\frac{\eta_2}{\eta_1}\;\frac{u_c}{u},\qquad u\;\frac{\textrm{d}}{\textrm{d}u}\bigg{[}u^m\;u_c^n \bigg{]}=\left( m+n\;\frac{\eta_2}{\eta_1} \right)u^m\;u_c^n,\\
&u^2\;\frac{\textrm{d}^2}{\textrm{d}u^2}\bigg{[}u^m\;u_c^n \bigg{]}=\left( m+n\;\frac{\eta_2}{\eta_1} \right)\left( m-1+n\;\frac{\eta_2}{\eta_1} \right)u^m\;u_c^n,
\end{split}
\right. \label{3:equ:derive:relation}
\end{eqnarray}
since $u_c$ is a function of $u$.  Besides, $u_c=1$ when $u=1$.  Using (\ref{2:equ:transformation}) and (\ref{2:equ:transformation:derive}), Eqs.~(\ref{2:equ:dimensionless:governing}) and (\ref{2:equ:dimensionless:boundary}) transform to
\begin{eqnarray}
{\cal N}_1\Big{[}w(u)\Big{]}&=&(A+1)\eta_1^2\left( u\;\frac{\textrm{d}w}{\textrm{d}u}+u^2\;\frac{\textrm{d}^2 w}{\textrm{d} u^2} \right)+w+8B\eta_1\; u \; \frac{\textrm{d}w}{\textrm{d} u}-\left(8B\eta_1+\eta_1^2\right) u\;w \;\frac{\textrm{d}w}{\textrm{d} u}\nonumber\\
&-&\eta_1^2 \; u^2\;w \;\frac{\textrm{d}^2w}{\textrm{d} u^2}-\frac{1}{2}\eta_1^2 \left(u\;\frac{\textrm{d}w}{\textrm{d} u}\right)^2=0, \label{2:equ:transformation:governing}
\end{eqnarray}
subjects to boundary conditions
\begin{equation}
w(u)\bigg{|}_{u=1}=1,\qquad \frac{\textrm{d}w(u)}{\textrm{d}u}\bigg{|}_{u=1}=0,\label{2:equ:transformation:boundary}
\end{equation}
where ${\cal N}_1$ is a nonlinear operator.

\section{The explicit series solution given by the HAM approach}
Assume that $w(u)$ can be expressed as
\begin{equation}
w(u)=\sum_{m+n\geq 1}a_{m,n}\;u^m\;u_c^n, \label{3:equ:solution:expression}
\end{equation}
where $a_{m,n}$ is a constant coefficient to be determined.  This provides us a so-called ``solution expression" of $w(u)$ in the frame of the HAM.  Besides, let $\gamma_0(u)$ denote the initial guess of $w(u)$, which satisfies the boundary conditions (\ref{2:equ:transformation:boundary}). Moreover, let $\cal L$ denote an auxiliary linear operator with the property ${\cal L}[0]=0$, $c_0$ a non-zero auxiliary parameter, called the convergence-control parameter, and $q\in [0,1]$ the embedding quantity, respectively.   Then we construct a family of differential equations
\begin{equation}
(1-q){\cal L}\Big{[}\Gamma(u;q)-\gamma_0(u)\Big{]}=c_0\;q\;{\cal N}_1\Big{[}\Gamma(u;q)\Big{]}, \label{3:equ:zero:deformation}
\end{equation}
subject to boundary conditions
\begin{equation}
\Gamma(u;q)\bigg{|}_{u=1}=1,\qquad \frac{\partial\Gamma(u;q)}{\partial u}\bigg{|}_{u=1}=0, \label{3:equ:zero:boundary}
\end{equation}
where the nonlinear operator ${\cal N}_1$ is defined by (\ref{2:equ:transformation:governing}).  Here $\Gamma(u;q)$ corresponds to the unknown function $w(u)$, as mentioned below.

When $q=0$, due to the property ${\cal L}[0]=0$, Eq.~(\ref{3:equ:zero:deformation}) has the solution
\begin{equation}
\Gamma(u;0)=\gamma_0(u). \label{3:equ:q0}
\end{equation}
When $q=1$, Eq.~(\ref{3:equ:zero:deformation}) is equivalent to original equation (\ref{2:equ:transformation:governing}), provided
\begin{equation}
\Gamma(u;1)=w(u). \label{3:equ:q1}
\end{equation}
Therefore, when the embedding quantity $q$ varies continuously from 0 to 1, $\Gamma(u;q)$ deforms from a given (known) initial guess $\gamma_0(u)$ to the unknown function $w(u)$.  Hence, Eqs.~(\ref{3:equ:zero:deformation}) and (\ref{3:equ:zero:boundary}) are called zeroth-order deformation equations.

Using (\ref{3:equ:q0}), $\Gamma(u;q)$ can be expanded into following Maclaurin series
\begin{equation}
\Gamma(u;q)=\sum_{m=0}^{+\infty}\gamma_m(u)\;q^m, \label{3:equ:Maclaurin:series}
\end{equation}
where
\begin{equation}
\gamma_m(u)={\cal D}_m\Big{[}\Gamma(u;q)\Big{]}
\end{equation}
in which
\begin{equation}
{\cal D}_m\big{[}f \big{]}=\frac{1}{m!}\frac{\partial^m f}{\partial q^m}\bigg{|}_{q=0} \label{3:equ:def:D}
\end{equation}
is called the $m$th-order homotopy-derivative of $f$.  Assume that the auxiliary linear operator $\cal L$ and the convergence-control parameter $c_0$ are so properly selected that the Maclaurin series (\ref{3:equ:Maclaurin:series}) is convergent at $q=1$, then according to (\ref{3:equ:q1}), we have the so-called homotopy-series solution
\begin{equation}
w(u)=\sum_{m=0}^{+\infty}\gamma_m(u).  \label{3:equ:homotopy:series:solution}
\end{equation}

Substituting (\ref{3:equ:Maclaurin:series}) into the zeroth-order deformation equations (\ref{3:equ:zero:deformation}), and then equating the coefficient of $q^m$ ($m\geq1$), we have the $m$th-order deformation equation
\begin{equation}
{\cal L}\Big{[}\gamma_m(u)-\chi_m \gamma_{m-1}(u) \Big{]}=c_0\;\delta_{m-1}(u), \label{3:equ:mth:governing}
\end{equation}
subjects to boundary conditions
\begin{equation}
\gamma_m(u)\bigg{|}_{u=1}=0,\qquad \frac{\textrm{d}\gamma_m(u)}{\textrm{d} u}\bigg{|}_{u=1}=0, \label{3:equ:mth:boundary}
\end{equation}
in which
\begin{eqnarray}
\begin{split}
\delta_m(u)&={\cal D}_m\Big{\{}{\cal N}_1\big{[}\Gamma(u;q)\big{]}  \Big{\}} \\
&=(A+1)\eta_1^2\left( u\;\frac{\textrm{d}\gamma_m}{\textrm{d}u}+u^2\;\frac{\textrm{d}^2 \gamma_m}{\textrm{d} u^2} \right)+\gamma_m+8B\eta_1 u \; \frac{\textrm{d}\gamma_m}{\textrm{d} u} \\
&-\sum_{i=0}^{m}\left[ \left(8B\eta_1+\eta_1^2\right)u\gamma_i \frac{\textrm{d}\gamma_{m-i}}{\textrm{d} u} + \eta_1^2\;u^2\;\gamma_i  \; \frac{\textrm{d}^2\gamma_{m-i}}{\textrm{d} u^2}+\frac{1}{2}\eta_1^2 \; u^2\;\frac{\textrm{d}\gamma_i}{\textrm{d} u}\frac{\textrm{d}\gamma_{m-i}}{\textrm{d} u}\right],\qquad \label{3:equ:delta}
\end{split}
\end{eqnarray}
and
\begin{equation}
\chi_m =\left\{
\begin{array}{cc}
0 & \mbox{when $m\leq 1$}, \\
1 & \mbox{when $m > 1$.}
\end{array}
\right. \label{3:equ:def:chi}
\end{equation}
Here ${\cal D}_m$ is defined by (\ref{3:equ:def:D}).  Note that $\gamma_0(u)$ should satisfy the boundary conditions (\ref{2:equ:transformation:boundary}).  According to the solution expression (\ref{3:equ:solution:expression}), we choose
\begin{equation}
\gamma_0(u)=\frac{\eta_1 }{\eta_1-\eta_2}u_c-\frac{\eta_2}{\eta_1-\eta_2}u \label{3:equ:initial:guess}
\end{equation}
as the initial guess of $w(u)$.  Besides, it is easy to find that there are no $u$ and $u_c$ in $\delta_m(u)$, i.e., $\delta_m(u)$ is made up of $u^m u_c^n$, in which $m+n>1$.  So we choose following auxiliary linear operator
\begin{equation}
{\cal L}[f]=\eta_1^2 \;u^2\;\frac{\textrm{d}^2 f}{\textrm{d}u^2}-\eta_1\eta_2\;u\;\frac{\textrm{d} f}{\textrm{d}u}+\eta_1\eta_2f. \label{3:equ:linear:operator}
\end{equation}
Then the corresponding inverse linear operator satisfies
\begin{equation}
{\cal L}^{-1}\Big{[}u^mu_c^n\Big{]}=\frac{u^m\;u_c^n}{\big{[}(m-1)\eta_1+n\eta_2\big{]} \big{[}m\eta_1+(n-1)\eta_2\big{]}}, \qquad m+n>1. \label{3:equ:inverse:linear:operator}
\end{equation}

Using the initial guess (\ref{3:equ:initial:guess}) and the inverse linear operator (\ref{3:equ:inverse:linear:operator}), the solution $\gamma_m(u)$ of the linear high-order deformation equations (\ref{3:equ:mth:governing}) and (\ref{3:equ:mth:boundary}) can be easily obtained step by step, starting from $m=1$, say,
\begin{equation}
\gamma_m(u)=\chi_m \gamma_{m-1}(u)+c_0 {\cal L}^{-1}\big{[}\delta_{m-1}(u) \big{]}+\Lambda_1^{(m)} u+\Lambda_2^{(m)} u_c,  \label{3:equ:general:solution}
\end{equation}
in which $\Lambda_1^{(m)}$ and $\Lambda_2^{(m)}$ are constant coefficients which are determined by the boundary conditions (\ref{3:equ:mth:boundary}).   Actually it is found that $\gamma_m(u)$ can be expressed as follows:
\begin{equation}
\gamma_m(u)=\sum_{i=0}^{m+1}\sum_{j=\max\{1-i,0\}}^{m+1-i}a_{i,j}^{(m)}u^i u_c^j, \label{3:equ:gamma:m:expression}
\end{equation}
with the recursion formula of $a_{i,j}^{(m)}$:
\begin{eqnarray}
\left\{
\begin{split}
&a_{0,1}^{(0)}=\frac{\eta_1}{\eta_1-\eta_2},\qquad a_{1,0}^{(0)}=\frac{\eta_2}{\eta_2-\eta_1},\\
&a_{i,j}^{(m+1)}=g_{i,j}^{(m+1)}+(1-\chi_i)(1-\chi_{j+1})\Lambda_1^{(m+1)}+(1-\chi_{i+1})(1-\chi_j)\Lambda_2^{(m+1)}, \label{3:equ:recursion:formula}
\end{split}
\right.
\end{eqnarray}
where
\begin{eqnarray}
&\;&\Lambda_1^{(m+1)}=\sum_{i=0}^{m+2}\sum_{j=\max\{1-i,0\}}^{m+2-i}\frac{g^{(m+1)}_{i,j}}{\eta_2-\eta_1}\Big{[}i\eta_1+(j-1)\eta_2\Big{]},\label{3:equ:recursion:Lambda1}\\
&\;&\Lambda_2^{(m+1)}=\sum_{i=0}^{m+2}\sum_{j=\max\{1-i,0\}}^{m+2-i}\frac{g^{(m+1)}_{i,j}}{\eta_2-\eta_1}\Big{[}(1-i)\eta_1-j\eta_2\Big{]}.\label{3:equ:recursion:Lambda2}\\
&\;&g_{i,j}^{(m+1)}=\chi_{m+1} \chi_{m+3-i-j}a_{i,j}^{(m)}+\frac{c_0\left[\Delta_{i,j}^{(1,m+1)}-\Delta_{i,j}^{(2,m+1)} \right]}{\big{[}(i-1)\eta_1+j\eta_2\big{]} \big{[}i\eta_1+(j-1)\eta_2\big{]}}.\label{3:equ:recursion:g}\\
&\;&\Delta_{i,j}^{(1,m+1)}=\chi_{m+3-i-j}\left\{(A+1)\eta_1^2\left[b_{i,j}^{(m)}+c_{i,j}^{(m)} \right]+a_{i,j}^{(m)}+8B\;\eta_1 \; b_{i,j}^{(m)}\right\},\label{3:equ:recursion:delta1}\\
&\;&\Delta_{i,j}^{(2,m+1)}=\sum_{n=0}^m \left[(8B\eta_1+\eta_1^2)\;\;d_{i,j}^{(m+1)}+\eta_1^2 \;e_{i,j}^{(m+1)}+\frac{1}{2}\;\eta_1^2\;f_{i,j}^{(m+1)}  \right].\label{3:equ:recursion:delta2}
\end{eqnarray}
in which
\begin{eqnarray}
&\;&b_{i,j}^{(n)}=a_{i,j}^{(n)}\left(i+j\;\frac{\eta_2}{\eta_1} \right),\quad\quad c_{i,j}^{(n)}=b_{i,j}^{(n)}\left(i-1+j\;\frac{\eta_2}{\eta_1} \right).\label{3:equ:recursion:b:c}\\
&\;&d_{i,j}^{(m+1)}=\sum_{p=\max\{0,1-j,i-m+n-1\}}^{\min\{i,n+1,i+j-1 \}}\;\sum_{r=\max\{0,1-p,i+j-m+n-1-p\}}^{\min\{j,n+1-p,i+j-1-p \}}a_{p,r}^{(n)}\;b_{i-p,j-r}^{(m-n)}. \label{3:equ:recursion:d}\\
&\;&e_{i,j}^{(m+1)}=\sum_{p=\max\{0,1-j,i-m+n-1\}}^{\min\{i,n+1,i+j-1 \}}\;\sum_{r=\max\{0,1-p,i+j-m+n-1-p\}}^{\min\{j,n+1-p,i+j-1-p \}}a_{p,r}^{(n)}\;c_{i-p,j-r}^{(m-n)},\label{3:equ:recursion:e}\\
&\;&f_{i,j}^{(m+1)}=\sum_{p=\max\{0,1-j,i-m+n-1\}}^{\min\{i,n+1,i+j-1 \}}\;\sum_{r=\max\{0,1-p,i+j-m+n-1-p\}}^{\min\{j,n+1-p,i+j-1-p \}}b_{p,r}^{(n)}\;b_{i-p,j-r}^{(m-n)}. \label{3:equ:recursion:f}
\end{eqnarray}

The $M$th-order homotopy-approximation of $w(u)$ reads
\begin{equation}
\tilde{\Gamma}_M(u)=\sum_{m=0}^{M}\gamma_m(u).
\end{equation}
As long as $\tilde{\Gamma}_M(u)$ is known, it is straightforward to gain the corresponding $M$th-order homotopy-approximation solution $\tilde{z}_M(\tau)$ of Eqs.~(\ref{2:equ:dimensionless:governing}) and (\ref{2:equ:dimensionless:boundary}) by using transformation (\ref{2:equ:transformation}).

In order to characterize the accuracy of the HAM, we define following squared residual error
\begin{equation}
{\cal \epsilon}=\int_0^{+\infty}\left\{{\cal N}\Big{[}\tilde{z}_M(\tau)\Big{]}   \right\}^2\textrm{d} \tau
\end{equation}
where $\cal N$ is defined by (\ref{2:equ:dimensionless:governing}).  Obviously, the smaller the squared residual error $\cal \epsilon$, the more accurate the HAM approximation.

\section{Results}
Without loss of generality, let us consider the liquid to be diethyl ether with property $\mu=2.2\cdot 10^{-4} \textrm{Pa}\cdot \textrm{s}$, $\sigma=1.67\cdot 10^{-2}\textrm{N} \cdot \textrm{m}^{-1}$, $\rho=710 \textrm{kg} \cdot\textrm{m}^{-3}$, the wetting angle $\theta=26^\circ$ and acceleration due to gravity $g=9.8 \textrm{m} \cdot\textrm{s}^{-2}$.  Note that according to (\ref{2:eta1:eta2}), the critical value reads $r_c\approx0.23\textrm{mm}$, corresponding to $16B^2-A-1=0$, above which the path of liquid in the capillary is oscillatory.  So we consider four cases $r=0.1\textrm{mm}$, $0.2\textrm{mm}$, $0.3\textrm{mm}$, $0.4\textrm{mm}$ respectively in this section so as to illustrate the validity of our explicit series solution given by the HAM for this problem.

\begin{figure}
\centering
{
\begin{minipage}[b]{2.5in}
\centerline{\includegraphics[width=2.2in]{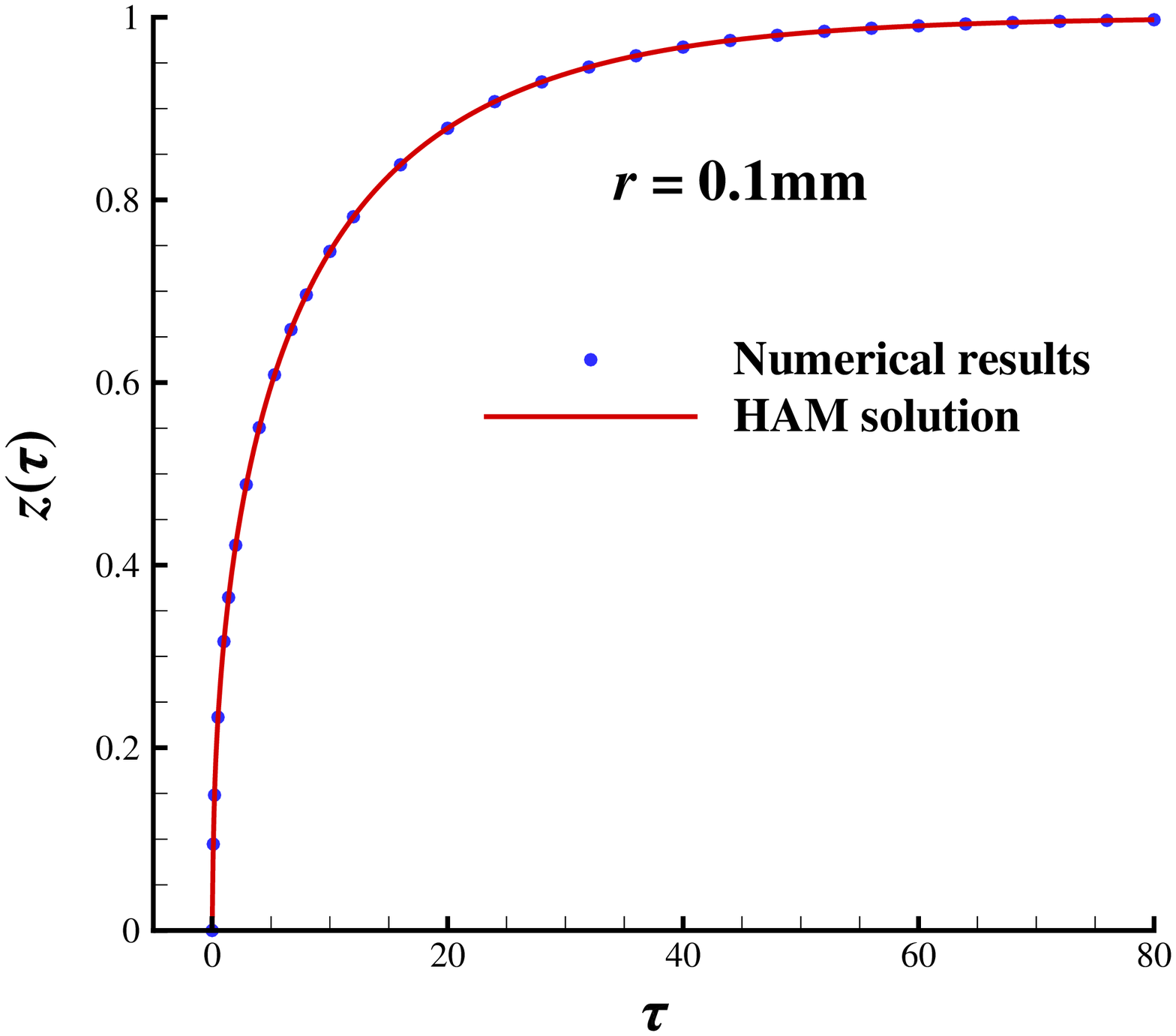}}
\centerline{(a)}
\end{minipage}
}
\hfill
{
\begin{minipage}[b]{2.5in}
\centerline{\includegraphics[width=2.2in]{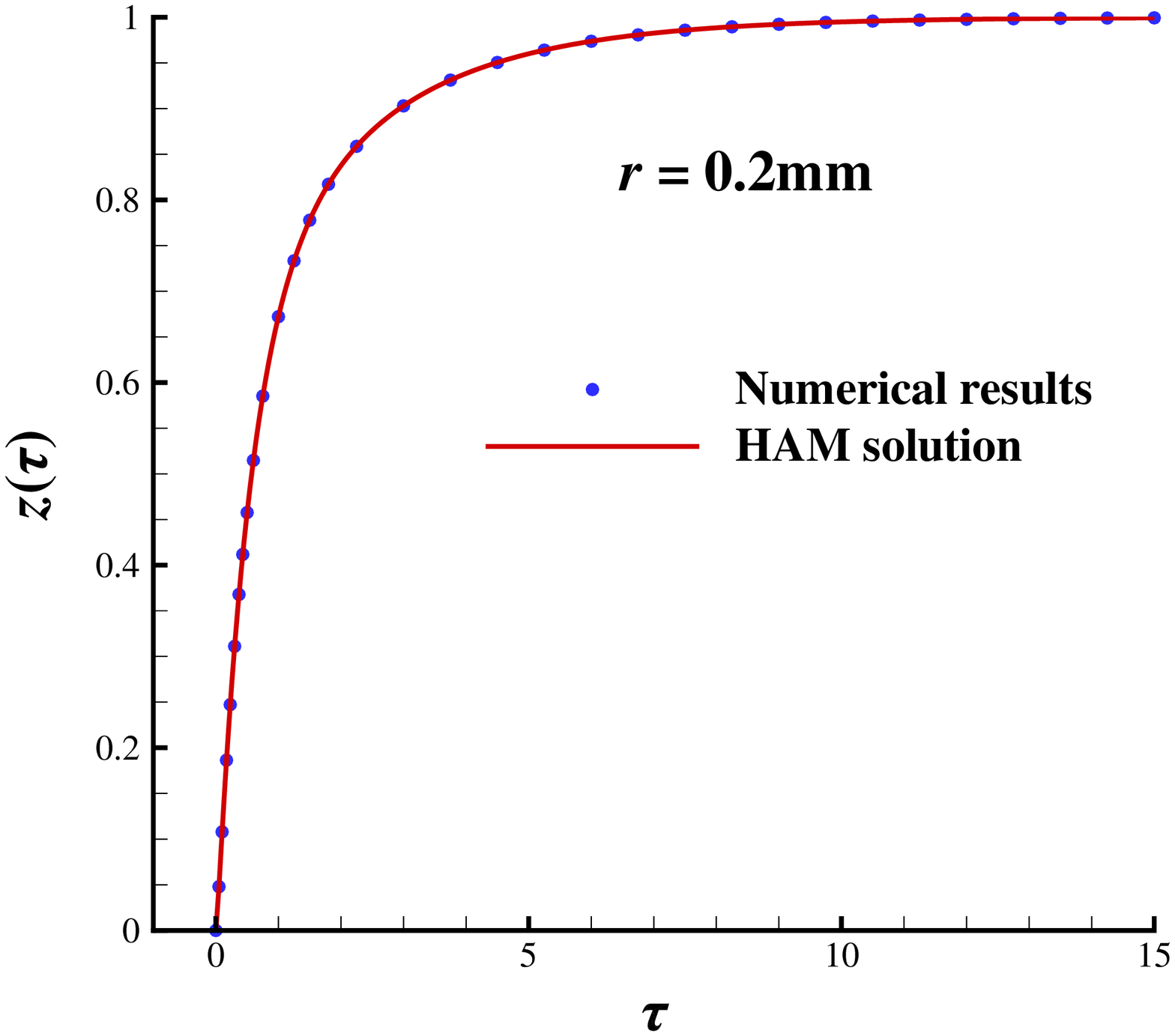}}
\centerline{(b)}
\end{minipage}
}
\vfill
{
\begin{minipage}[b]{2.5in}
\centerline{\includegraphics[width=2.2in]{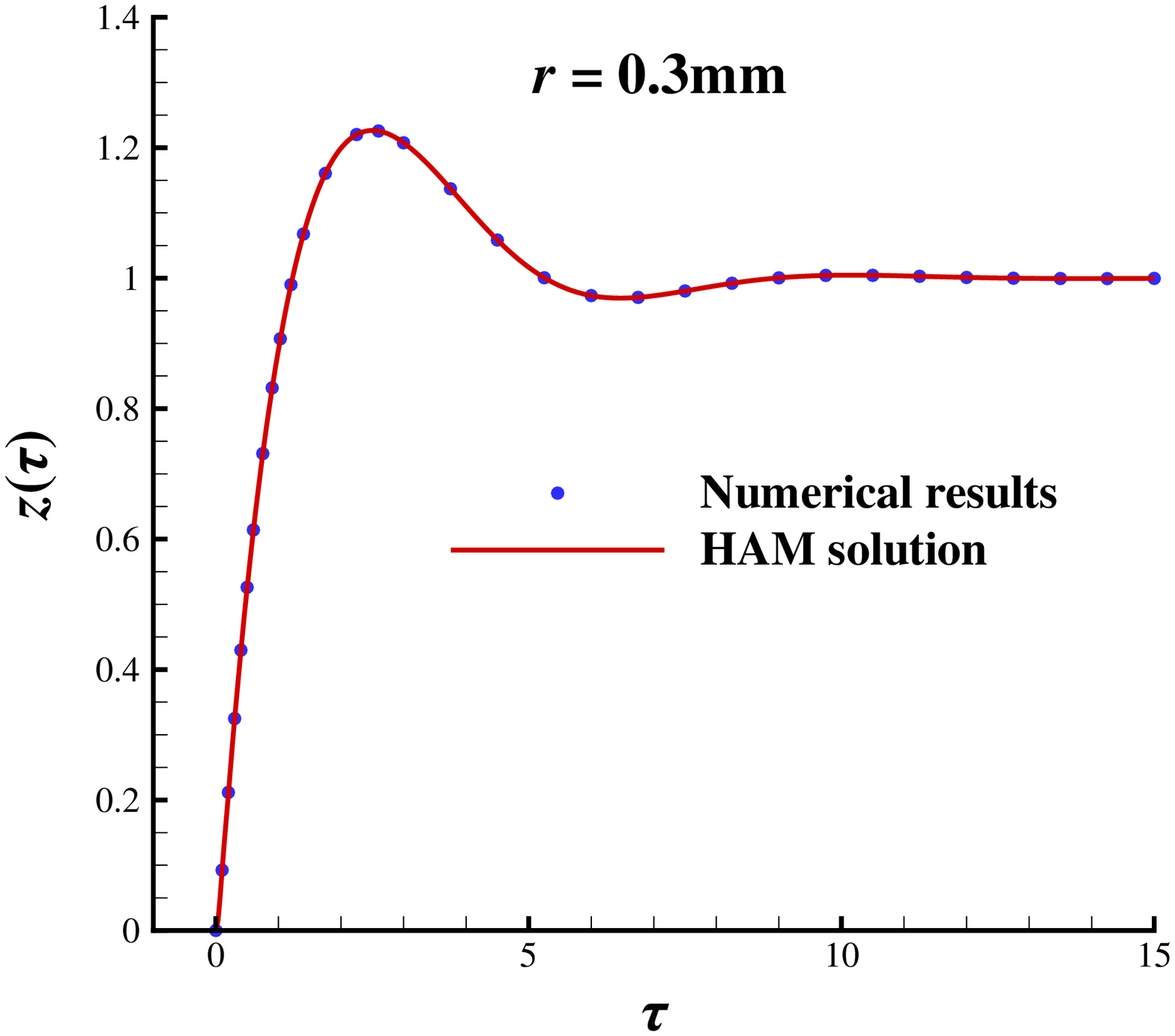}}
\centerline{(c)}
\end{minipage}
}
\hfill
{
\begin{minipage}[b]{2.5in}
\centerline{\includegraphics[width=2.2in]{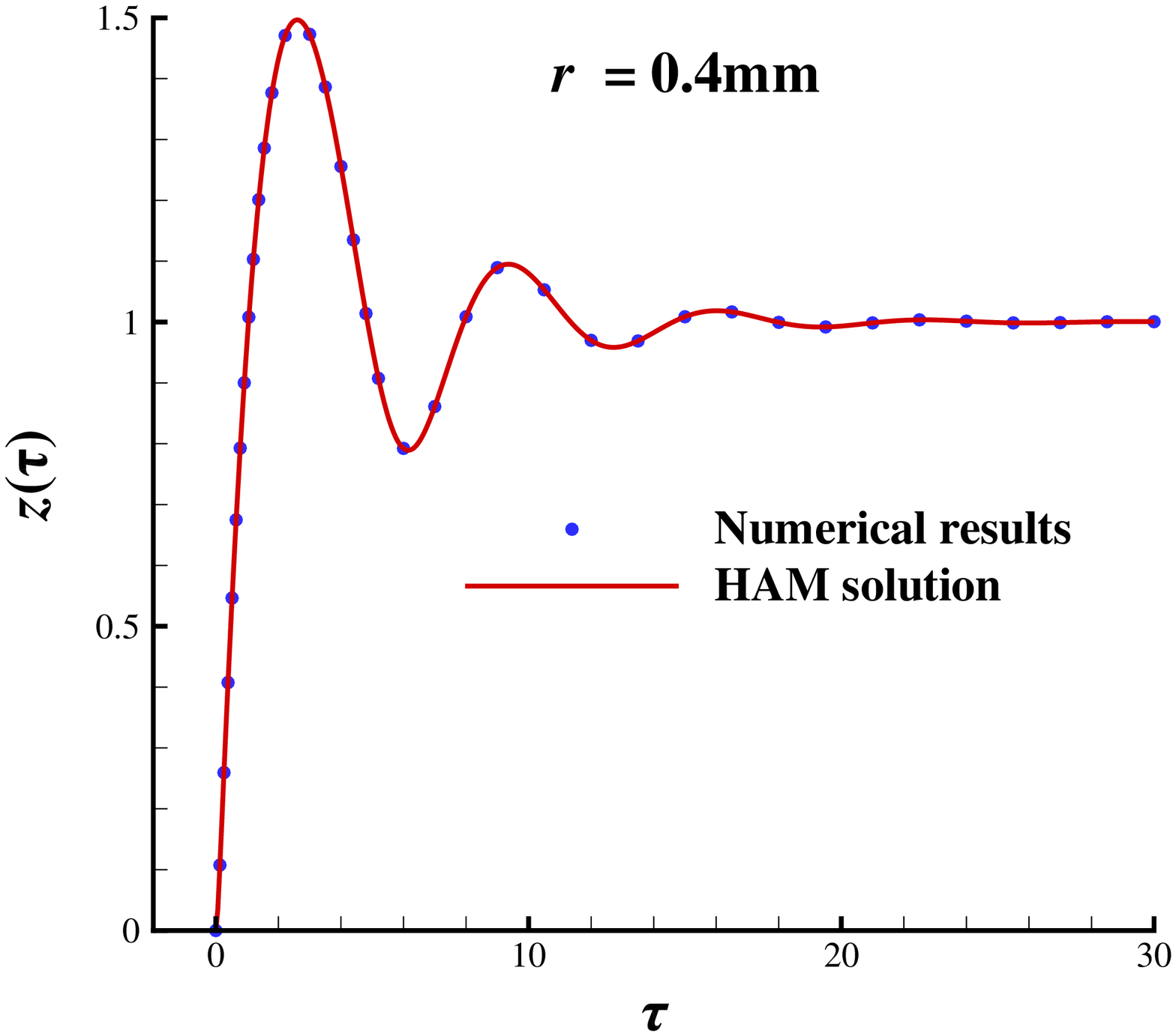}}
\centerline{(d)}
\end{minipage}
}
\caption{Comparison between the homotopy-series solution given by the HAM and the numerical results given by the symbolic software Mathematica using six-order Runge-Kutta methods.  (a) $r=0.1$mm; (b) $r=0.2$mm; (c) $r=0.3$mm; (d) $r=0.4$mm.}
\label{comparison:homotopy:numerical}
\end{figure}

Note that there is an unknown auxiliary convergence-control parameter $c_0$ in (\ref{3:equ:general:solution}), which provides us a convenient way to guarantee the convergence of homotopy-series solution (\ref{3:equ:homotopy:series:solution}).  It is found that different cases share the same convergent region $c_0\in [-1,0)$, and the same optimal convergence-control parameter $c_0=-1$ (corresponding to the minimum squared residual error $\cal \epsilon$).  Using $c_0=-1$, convergent results can be easily obtained, as shown in Table \ref{table:a:1}-\ref{table:a:4}.  Here, it is worth mentioning that the smaller the radius of circular tube $r$ is, the slower the rate of convergence is, as shown in Table \ref{table:a:1}-\ref{table:a:4}.  This is understandable since for a rather thin tube, corresponding to a small $r$, the acceleration at the initial stage will be very large and hence the behavior of liquid will change a lot in a pretty short time.  In order to well describe this stage, higher order approximations are therefore needed.  The comparisons between the homotopy-series solution and the numerical results given by the symbolic software Mathematica using six-order Runge-Kutta methods are as shown in Fig.~\ref{comparison:homotopy:numerical}.  It is found that our results agree very well with numerical results no matter the path of liquid in a circular tube approaching the final equilibrium position is monotonic or oscillatory.  This illustrates that our explicit series solution is convergent and valid for all cases.

\begin{table}
\tabcolsep 0pt
\begin{center}
\def\temptablewidth{0.9\textwidth}
\begin{tabular*}{\temptablewidth}{@{\extracolsep{\fill}}cccccc}
$m$, order of approx. &${\cal E}_m$ & $z_m(1/4)$    &  $z_m(1)$ &$z_m(5)$ & $z_m(20) $     \\
1 & $1.1\times10^{-1}$ &0.023   & 0.109   &  0.455    &   0.914          \\
30 & $8.9\times10^{-5}$ &0.175   & 0.320  &  0.596     &   0.878      \\
60 & $6.0\times10^{-5}$ &0.165   & 0.315   &  0.596    &   0.878      \\
90 &$2.5\times10^{-5}$ &0.165   & 0.316   &    0.596    &   0.878            \\
120 & $5.6\times10^{-6}$ & 0.168   & 0.316 &  0.596      &   0.878                     \\
150 & $4.3\times10^{-7}$&0.167 & 0.316   &  0.596      &   0.878                       \\
180 & $3.5\times10^{-8}$&0.167   & 0.316  &   0.596     &   0.878                      \\
\end{tabular*}
\caption{Analytic approximations of $z(\tau)$ in Eq.~(\ref{2:equ:dimensionless:governing}) in the case of $r=0.1\textrm{mm}$, given by the HAM using the convergence-control parameter $c_0=-1$.}   \label{table:a:1}
 \end{center}
 \end{table}

  \begin{table}
 \tabcolsep 0pt
\begin{center}
\def\temptablewidth{0.9\textwidth}
\begin{tabular*}{\temptablewidth}{@{\extracolsep{\fill}}cccccc}
$m$, order of approx. &${\cal E}_m$ & $z_m(1/4)$    &  $z_m(1/2)$ &$z_m(1)$ & $z_m(2) $     \\
1 & $1.5\times10^{-3}$ &0.049   & 0.155   &  0.403    &   0.759          \\
30 & $9.7\times10^{-4}$ &0.260   & 0.457  &  0.669     &   0.837      \\
60 & $8.9\times10^{-5}$ &0.265   & 0.457   &  0.672    &   0.837      \\
90 &$5.8\times10^{-6}$ &0.266   & 0.457   &    0.672    &   0.837            \\
120 & $7.9\times10^{-7}$ & 0.266   & 0.457 &  0.672      &   0.838                     \\
150 & $1.3\times10^{-7}$&0.266 & 0.458   &  0.672      &   0.838                       \\
180 & $3.2\times10^{-8}$&0.266   & 0.458  &   0.672     &   0.838                      \\
\end{tabular*}
\caption{Analytic approximations of $z(\tau)$ in Eq.~(\ref{2:equ:dimensionless:governing}) in the case of $r=0.2\textrm{mm}$, given by the HAM using the convergence-control parameter $c_0=-1$.}   \label{table:a:2}
 \end{center}
 \end{table}

  \begin{table}
 \tabcolsep 0pt
\begin{center}
\def\temptablewidth{1\textwidth}
\begin{tabular*}{\temptablewidth}{@{\extracolsep{\fill}}ccccccc}
$m$, order of approx. &$\cal E$& $z_m(1/4)$    &  $z_m(1/2)$ &$z_m(1)$ & $z_m(2) $  & $z_m(4) $    \\
1 &$2.1\times 10^{-3}$&0.0555   & 0.1979   &  0.6002    &   1.2124 &1.2248          \\
40 &$3.3\times 10^{-4}$& 0.2673   & 0.5257  &  0.8897     &   1.1992    &   1.1101    \\
80 &$4.7\times 10^{-5}$& 0.2691   & 0.5259  &  0.8905     &   1.1994    &   1.1099    \\
120 &$1.9\times 10^{-6}$ & 0.2691   & 0.5260  &  0.8906     &   1.1994    &   1.1099    \\
160& $2.0\times 10^{-7}$&0.2692   & 0.5260  &  0.8906     &   1.1994    &   1.1099    \\
200 &$8.7\times 10^{-9}$& 0.2692   & 0.5260  &  0.8906     &   1.1994    &   1.1099    \\
\end{tabular*}
\caption{Analytic approximations of $z(\tau)$ in Eq.~(\ref{2:equ:dimensionless:governing}) in the case of $r=0.3\textrm{mm}$, given by the HAM using the convergence-control parameter $c_0=-1$.}   \label{table:a:3}
 \end{center}
 \end{table}

   \begin{table}
 \tabcolsep 0pt
\begin{center}
\def\temptablewidth{1\textwidth}
\begin{tabular*}{\temptablewidth}{@{\extracolsep{\fill}}ccccccc}
$m$, order of approx. &$\cal E$& $z_m(1/4)$    &  $z_m(1/2)$ &$z_m(1)$ & $z_m(2) $  & $z_m(4) $    \\
1 &$1.1\times 10^{-2}$&0.0569   & 0.2111   &  0.6831    &   1.4658 &1.3014          \\
40 &$1.5\times 10^{-3}$& 0.2474   & 0.5254  &  0.9682     &   1.4319    &   1.2559    \\
80 &$1.7\times 10^{-5}$& 0.2479   & 0.5257  &  0.9684     &   1.4320    &   1.2557    \\
120 &$1.2\times 10^{-7}$ & 0.2479   & 0.5257  &  0.9684     &   1.4320    &   1.2557    \\
160& $2.7\times 10^{-8}$&0.2479   & 0.5257  &  0.9684     &   1.4320    &   1.2557    \\
200 &$7.6\times 10^{-10}$&0.2479   & 0.5257  &  0.9684     &   1.4320    &   1.2557    \\
\end{tabular*}
\caption{Analytic approximations of $z(\tau)$ in Eq.~(\ref{2:equ:dimensionless:governing}) in the case of $r=0.4\textrm{mm}$, given by the HAM using the convergence-control parameter $c_0=-1$.}   \label{table:a:4}
 \end{center}
 \end{table}

In the case of $r=0.1$mm, corresponding to the monotonic path of liquid in a capillary, the homotopy solution series $\gamma_m$ reads
\begin{eqnarray}
\left\{
\begin{split}
&\gamma_0=1.004\textrm{e}^{-0.0611\tau}-0.003756\textrm{e}^{-16.325\tau},\\
&\gamma_1=-1.012\textrm{e}^{-0.0611\tau}+1.016\textrm{e}^{-0.1222\tau}-0.01037\textrm{e}^{-16.325\tau}+0.006518\textrm{e}^{-16.386\tau}\\
&\qquad +3.488\cdot 10^{-6}\textrm{e}^{-32.651\tau},\\
&\gamma_2=0.5121\textrm{e}^{-0.0611\tau}-2.052\textrm{e}^{-0.1222\tau}+1.544\textrm{e}^{-0.1833\tau}-0.01626\textrm{e}^{-16.325\tau}\\
&\qquad+0.01140\textrm{e}^{-16.386\tau}+9.054\cdot 10^{-4}\textrm{e}^{-16.448\tau}+1.924\cdot 10^{-5}\textrm{e}^{-32.651\tau}\\
&\qquad-8.657\cdot 10^{-6}\textrm{e}^{-32.712\tau}-8.702\cdot 10^{-9}\textrm{e}^{-48.976\tau},\\
&\gamma_3=-0.1740\textrm{e}^{-0.0611\tau}+2.075\textrm{e}^{-0.1222\tau}-4.678\textrm{e}^{-0.1833\tau}+2.780\textrm{e}^{-0.2444\tau}\\
&\qquad
-0.01994\textrm{e}^{-16.325\tau}+0.01338\textrm{e}^{-16.386\tau}+6.678\cdot 10^{-4}\textrm{e}^{-16.448\tau}\\
&\qquad +1.792\cdot 10^{-3}\textrm{e}^{-16.509\tau}+5.671\cdot 10^{-5}\textrm{e}^{-32.651\tau}-3.900\cdot 10^{-5}\textrm{e}^{-32.712\tau}\\
&\qquad +3.739\cdot 10^{-6}\textrm{e}^{-32.773\tau}-7.200\cdot 10^{-8}\textrm{e}^{-48.976\tau}+2.801\cdot 10^{-8}\textrm{e}^{-49.073\tau}\\
&\qquad+2.851\cdot 10^{-11}\textrm{e}^{-65.302\tau},\\
&\cdots
\end{split}  \label{equ:4:solution:case:1}
\right.
\end{eqnarray}
By contrast, in the case of $r=0.4$mm, corresponding to the oscillatory path of liquid in a capillary, the homotopy solution series $\gamma_m$ reads
\begin{eqnarray}
\left\{
\begin{split}
&\gamma_0=\textrm{e}^{-0.246\tau}\Big{[}\cos(0.948\tau)+0.260\sin(0.948\tau)\Big{]},\\
&\gamma_1=\textrm{e}^{-0.246\tau}\Big{[}0.0647\cos(0.948\tau)-0.359\sin(0.948\tau)\Big{]}+\textrm{e}^{-0.492\tau}\Big{[}-0.270\\
&\qquad+0.205\cos(1.895\tau)+0.171\sin(1.895\tau)\Big{]},\\
&\gamma_2=\textrm{e}^{-0.246\tau}\Big{[}-0.0509\cos(0.948\tau)-0.093\sin(0.948\tau)\Big{]}+\textrm{e}^{-0.492\tau}\Big{[}0.0262\\
&\qquad +0.1008\cos(1.895\tau)-0.161\sin(1.895\tau)\Big{]}+\textrm{e}^{-0.738\tau}\Big{[}-0.136\cos(0.948\tau)\\
&\qquad +0.0967\sin(0.948\tau)+0.0595\cos(2.843\tau)+0.104\sin(2.843\tau)\Big{]},\\
&\gamma_3=\textrm{e}^{-0.246\tau}\Big{[}-0.0272\cos(0.948\tau)-0.0240\sin(0.948\tau)\Big{]}+\textrm{e}^{-0.492\tau}\Big{[}0.0031\\
&\qquad -0.0362\cos(1.895\tau)-0.065\sin(1.895\tau)\Big{]}+\textrm{e}^{-0.738\tau}\Big{[}0.057\cos(0.948\tau)\\
&\qquad +0.0316\sin(0.948\tau)+0.100\cos(2.843\tau)-0.080\sin(2.843\tau)\Big{]}\\
&\qquad +\textrm{e}^{-0.984\tau}\Big{[}-0.0339-0.075\cos(1.895\tau)+0.048\sin(1.895\tau)\\
&\qquad +0.0122\cos(3.790\tau)+0.066\sin(3.790\tau)\Big{]},\\
&\cdots
\end{split}  \label{equ:4:solution:case:4}
\right.
\end{eqnarray}

\section{Concluding remarks}
In this paper, we applied an analytic approximation method, namely the homotopy analysis method (HAM), to the rise dynamics of a liquid in a vertical circular tube.  By means of the HAM, explicit series solution is successfully obtained which is valid for all cases no matter the path of liquid is monotonic or oscillatory.  To the best knowledge of the authors, this kind of explicit series solution has never been reported previously.  This further demonstrates the validity and superiority of the HAM over other traditional analytic approximation methods.

Here, it should be emphasized that unlike perturbation method, the HAM is independent of \emph{any} small/large physical parameter.  More importantly, the HAM provides us great freedom to choose appropriate basis functions, the auxiliary linear operator $\cal L$ and the convergence-control parameter $c_0$.  It is this kind of freedom that allows us to choose such basis function (\ref{3:equ:solution:expression}) which can describe both monotonic and oscillatory path, as shown in (\ref{equ:4:solution:case:1}) and (\ref{equ:4:solution:case:4}).  As a result, our explicit series solution is valid for all cases whether the path of liquid is monotonic or not.  Besides, our approach is very simple to use.  All of these might deepen our understandings and enrich our knowledge about capillarity.

\section*{Acknowledgement}  This work is partly supported by National Natural Science Foundation of China (Approval No. 11432009).

\appendix
\section{The recursion formula of $a_{i,j}^{(m)}$ in (\ref{3:equ:gamma:m:expression})}
Here, we prove that $\gamma_m(u)$ can be expressed as (\ref{3:equ:gamma:m:expression}) and also derive the recursion formula of $a_{i,j}^{(m)}$ in (\ref{3:equ:recursion:formula}).

According to (\ref{3:equ:initial:guess}) and (\ref{3:equ:gamma:m:expression}), it is easy to gain $a_{0,1}^{(0)}$ and $a_{1,0}^{(0)}$ in (\ref{3:equ:recursion:formula}).
Assume that for any $n\leq m$, $\gamma_n(u)$ can be expressed as
\begin{equation}
\gamma_n(u)=\sum_{i=0}^{n+1}\sum_{j=\max\{1-i,0\}}^{n+1-i}a_{i,j}^{(n)}u^i u_c^j=\sum_{i=0}^{n+2}\sum_{j=\max\{1-i,0\}}^{n+2-i}\chi_{n+3-i-j}\;a_{i,j}^{(n)}u^i u_c^j. \label{appendix:gamma:expression}
\end{equation}
where $\chi_n$ is defined by (\ref{3:equ:def:chi}).  Then using (\ref{3:equ:derive:relation}), for any $n\leq m$, it holds
\begin{eqnarray}
\left\{
\begin{split}
u\frac{\textrm{d}\gamma_n}{\textrm{d}u}&=\sum_{i=0}^{n+1}\sum_{j=\max\{1-i,0\}}^{n+1-i}b_{i,j}^{(n)}u^i u_c^j=\sum_{i=0}^{n+2}\sum_{j=\max\{1-i,0\}}^{n+2-i}\chi_{n+3-i-j}\;b_{i,j}^{(n)}u^i u_c^j,\\
u^2\frac{\textrm{d}^2\gamma_n}{\textrm{d}u}&=\sum_{i=0}^{n+1}\sum_{j=\max\{1-i,0\}}^{n+1-i}c_{i,j}^{(n)}u^i u_c^j=\sum_{i=0}^{n+2}\sum_{j=\max\{1-i,0\}}^{n+2-i}\chi_{n+3-i-j}\;c_{i,j}^{(n)}u^i u_c^j,
\end{split} \label{appendix:gamma:derivative:expression}
\right.
\end{eqnarray}
in which $b_{i,j}^{(n)}$ and $c_{i,j}^{(n)}$ are defined by (\ref{3:equ:recursion:b:c}).

Using (\ref{appendix:gamma:expression}) and (\ref{appendix:gamma:derivative:expression}), it is easy to derive
\begin{eqnarray}
\begin{split}
&\quad\gamma_n\left(u\frac{\textrm{d}\gamma_{m-n}}{\textrm{d}u}\right)\\
&=\left(\sum_{p=0}^{n+1}\sum_{r=\max\{1-p,0\}}^{n+1-p}a_{p,r}^{(n)}u^p u_c^r\right)\left(\sum_{s=0}^{m-n+1}\sum_{t=\max\{1-s,0\}}^{m-n+1-s}b_{s,t}^{(m-n)}u^s u_c^t\right)\\
&=\sum_{i=0}^{m+2}\sum_{j=\max\{1-i,0\}}^{m+2-i}d_{i,j}^{(m+1)}u^i u_c^j,
\end{split}
\end{eqnarray}
where $d_{i,j}^{(m+1)}$ is defined by (\ref{3:equ:recursion:d}).  Similarly, we have
\begin{eqnarray}
\begin{split}
\gamma_n\left(u^2\frac{\textrm{d}^2\gamma_{m-n}}{\textrm{d}u^2}\right)&=\sum_{i=0}^{m+2}\sum_{j=\max\{1-i,0\}}^{m+2-i}e_{i,j}^{(m+1)}u^i u_c^j, \\
\left(u\frac{\textrm{d}\gamma_{n}}{\textrm{d}u}\right)\left(u\frac{\textrm{d}\gamma_{m-n}}{\textrm{d}u}\right)&=\sum_{i=0}^{m+2}\sum_{j=\max\{1-i,0\}}^{m+2-i}f_{i,j}^{(m+1)}u^i u_c^j,
\end{split} \label{appendix:e:f:derive}
\end{eqnarray}
where $e_{i,j}^{(m+1)}$ and $f_{i,j}^{(m+1)}$ are defined by (\ref{3:equ:recursion:e}) and (\ref{3:equ:recursion:f}) respectively.
Then according to (\ref{3:equ:delta}) and (\ref{appendix:gamma:expression})-(\ref{appendix:e:f:derive}), we gain
\begin{eqnarray}
\delta_m=\sum_{i=0}^{m+2}\sum_{j=\max\{1-i,0\}}^{m+2-i}\left[\Delta_{i,j}^{(1,m+1)}-\Delta_{i,j}^{(2,m+1)} \right]u^i u_c^j, \label{appendix:delta}
\end{eqnarray}
where $\Delta_{i,j}^{(1,m+1)}$ and $\Delta_{i,j}^{(2,m+1)}$ are defined by (\ref{3:equ:recursion:delta1}) and (\ref{3:equ:recursion:delta2}).

Using (\ref{3:equ:inverse:linear:operator}), (\ref{3:equ:general:solution}), (\ref{appendix:gamma:expression}) and (\ref{appendix:delta}), we have
\begin{eqnarray}
\begin{split}
\gamma_{m+1}&=\chi_{m+1}\gamma_m+c_0{\cal L}^{-1}[\delta_m]+\Lambda_1^{(m+1)}u+\Lambda_2^{(m+1)}u_c\\
&=\left(\sum_{i=0}^{m+2}\sum_{j=\max\{1-i,0\}}^{m+2-i}g_{i,j}^{(m+1)}u^i u_c^{j}\right)+\Lambda_1^{(m+1)}u+\Lambda_2^{(m+1)}u_c,
\end{split}
\end{eqnarray}
in which $g_{i,j}^{(m+1)}$ is defined by (\ref{3:equ:recursion:g}).  Then using boundary conditions (\ref{3:equ:mth:boundary}), it is easy to gain the expression of $\Lambda_1^{(m+1)}$ and $\Lambda_2^{(m+1)}$, as shown in (\ref{3:equ:recursion:Lambda1}) and (\ref{3:equ:recursion:Lambda2}).
Note that $u$ and $u_c$ can also be expressed as
\begin{eqnarray}
\left\{
\begin{split}
&u=\sum_{i=0}^{m+2}\sum_{j=\max\{1-i,0\}}^{m+2-i}\big{(}1-\chi_i\big{)}\big{(}1-\chi_{j+1}\big{)}u^i u_c^j,\\
&u_c=\sum_{i=0}^{m+2}\sum_{j=\max\{1-i,0\}}^{m+2-i}\big{(}1-\chi_{i+1}\big{)}\big{(}1-\chi_j\big{)}u^i u_c^j.
\end{split}
\right.
\end{eqnarray}
Therefore, we have
\begin{equation}
\gamma_{m+1}=\sum_{i=0}^{m+2}\sum_{j=\max\{1-i,0\}}^{m+2-i}a_{i,j}^{(m+1)}u^i u_c^{j},
\end{equation}
in which $a_{i,j}^{(m+1)}$ is defined by (\ref{3:equ:recursion:formula}).

\bibliographystyle{jfm}
\bibliography{reference}

\end{document}